\documentclass{elsart}

\usepackage{natbib}

\usepackage{epsfig}

\usepackage{amssymb}

\begin{document}

\newcommand{\INT}{{\it INTEGRAL}\,}
\newcommand{\XMM}{{\it XMM-Newton}\,}
\def\uu{4U\,0142+614\,}  
\def\ee{1E\,1048-5937\,} 
\def\kes{1E\,1841-045\,}
\def\axj{AX\,J1844-0258\,}
\def\rxs{1RXS\,J1708-4009\,}
\def\ea{1E\,2259+586\,}
\def\xte{XTE\,J1810-197\,}
\def\cxo{CXOU\,J0100-7211\,}
\def\wes{CXOU\,J1647-4552\,}
\def\1e{1E\,1547.0-5408\,}
\def\sgra{SGR\,1806-20\,}
\def\sgrb{SGR\,1900+14\,}
\def\sgrc{SGR\,0526-66\,}
\def\sgrd{SGR\,1627-41\,}
\def\sgre{SGR\,1801-23\,}
\def\sgrf{SGR\,0501+4516\,}

\begin{frontmatter}

\title{Modeling the broadband persistent emission of 
magnetars\thanksref{footnote1}}
\thanks[footnote1]{This template can be used for all publications in Advances in Space Research.}

\author{Silvia Zane\corauthref{cor}}
\address{Mullard Space Science Laboratory, University College of 
London, Holmbury St Mary Dorking, Surrey, RH5 6NT,  UK}
\corauth[cor]{Corresponding author}
\ead{sz@mssl.ucl.ac.uk}

\author{Roberto Turolla and Luciano Nobili}
\address{Dept. of Physics, University of Padova, via Marzolo 8, 
35131, Padova, 
Italy}
\ead{roberto.turolla@pd.infn.it, luciano.nobili@pd.infn.it}

\author{Nanda Rea}
\address{Astronomical Institute ``Anton Pannekoek'',  University of
Amsterdam, Science Park 904, Postbus 94249,
1090 GE, Amsterdam, NL}
\ead{N.Rea@uva.nl}

\begin{abstract} 
In this paper, we discuss our first attempts to model the 
broadband 
persistent emission of magnetars within a self consistent, physical 
scenario. We present the predictions of a synthetic model that we 
calculated with a new Monte Carlo 3-D radiative code. The basic idea is 
that soft thermal photons (e.g. emitted by the 
star surface) can experience resonant cyclotron upscattering 
by a population of relativistic electrons threated in the 
twisted magnetosphere. Our code is specifically tailored to work in 
the ultra-magnetized regime; polarization and QED effects are 
consistently accounted for, 
as well different configurations for the magnetosphere. We 
discuss the predicted spectral properties in the 
$0.1-1000$~keV range, the polarization properties, and we  
present the model application to a sample of magnetars 
soft X-ray spectra.
\end{abstract}


\end{frontmatter}

\parindent=0.5 cm

\section{Introduction}
Soft gamma-ray repeaters (SGRs) and anomalous X-ray pulsars (AXPs) are
peculiar X-ray sources which are believed to be magnetars:
ultra-magnetized neutron stars with surface field in excess of
$10^{14}$~G, i.e. well above the QED threshold \citep[see][for a
recent review]{sandrorev}. In this scenario, the ultra-strong 
field is responsible for powering the X-ray emission of the source which 
otherwise is too high to be explained in terms of spin down losses. 
Spectral analysis is an important tool in magnetar  
astrophysics since it can provide key information on the emission
mechanisms. The persistent (i.e. outside bursts) soft X-ray ($<10$~keV)
spectra of magnetars is typically reproduced by a double component model,
consisting of a blackbody (BB, with $kT \sim 0.3-0.6$~keV) plus a
power-law (PL, with photon
index $\Gamma \sim 2-4$). Moreover, \INT\, observations have shown that, while in
quiescence, magnetars emit substantial persistent radiation also at higher
energies, up to a few hundreds of keV. The X-ray persistent emission
above 20 keV has a power-law spectral shape ($\Gamma_h\sim 2$) which, in
particular in AXPs, is markedly harder than that observed below 10 keV
\citep[see again][and references therein]{sandrorev}. Despite these
phenomenological fitting models have been applied for many
years, a physical interpretation of the various spectral components is
still missing.

It has been proposed by several authors 
\citep{dt92, td93, tlk02} that, at variance with standard radio-pulsars,  
in magnetars the magnetosphere can be  twisted. This, in turn, has a 
number of observational consequences. 
Twisted magnetic fields are permeated by currents with  
substantial density, much in excess of the Goldreich-Julian current which
is expected in a potential field. Magnetospheric charges can   
efficiently interact with the primary radiation from the neutron star 
and/or with the magnetic field itself, giving rise
to different emission processes.
In particular, it has been widely suggested that the BB+PL spectral shape
that is observed
below $\sim10$~keV may be accounted for if the soft,
thermal spectrum
emitted by the star surface is distorted by resonant cyclotron scattering
(RCS) onto the magnetospheric charges. Since electrons permeate a
spatially extended region of the magnetosphere, where the magnetic field
varies by order of magnitudes, resonant scattering is not expected to 
manifest itself as a series of narrow spectral lines (corresponding to
the successive harmonics), but instead to lead to the
formation of a hard tail superimposed to the seed thermal bump.

\section{Building spectral models}

Recently, several efforts have been
carried out in order to test the resonant cyclotron scattering
model  quantitatively against
real data in the soft X-ray range, using different approaches and with a
varying degree of
sophistication. 
The first, seminal attempts have been presented by \citet{lg06}. These
authors studied  a simplified, 1-D model by assuming that
seed photons are
emitted by the NS surface radially and with a blackbody spectrum, and
magnetic Thomson
scattering occurs in a thin, plane parallel magnetospheric slab
permeated by a static, non-relativistic, warm medium at constant
electron density. They neglect all effects of electron recoil, as those
related to the currents bulk motion. Despite the simplification, this
model
has the main advantage to be semi-analytical and, when systematically
applied to X-ray data, has been proved to be
successful in catching the gross characteristics of the observed
soft X-ray spectrum \citep{nanda1,nanda3,nanda2}. The same model has been
extended by
\citet{guv07} who relaxed the BB approximation for the surface radiation  
and included atmospheric effects. More
recently, 3-D Monte Carlo calculations have been presented by
\citet{ft07}, although these spectra have never been applied to fit   
X-ray observations.
In this paper, we present the predicted spectra obtained via a new 3-D
Monte Carlo simulation \citep[see][for details]{ntz1,ntz2}. 
Polarization and, for the first time,  QED effects are consistently
accounted
for, as well different configurations for the magnetosphere. We will   
discuss the predicted spectral properties in the
in the   $0.1-1000$~keV range, the polarization properties
and we will present the model application
to a sample of magnetars spectra in the soft X-ray range.

\section{A Monte Carlo Model}

We decided to use a Monte Carlo approach since this  technique is 
particularly suitable for the problem at hand for several reasons. 
For instance, it allows us
to follow
individually a large sample of photons, treating probabilistically
their interactions with charged particles; it can handle very general 
(3-D)
geometries; is quite easy to code, and relatively fast;
and is ideal for purely scattering media.
In order to perform such kind of simulations, we need to specify
three basic ingredients: a) the space and energy distribution of seed
photons, b) the same for the magnetospheric electrons (the ``scatterers'')
and c) a prescription for the cross section. Although our code is
completely general, and can handle different thermal maps, different
models of surface emission  and magnetic
configurations, the models presented in this draft  are all computed
by assuming that the whole surface emits isotropically as a BB at a
single temperature, $kT$, and
that the magnetic field is a force-free, self-similar, twisted 
dipole characterized by the value of the twist angle, $\Delta \Phi$ and of
the polar magnetic field $B$ \citep{tlk02}.
In absence of a self-consistent treatment of the
electrodynamics of the magnetosphere \citep[see][]{luciano}, 
the electron
velocity distribution has been assumed a priori.
Our model
is based on a simplified treatment of the charge carriers velocity
distribution which accounts for the particle collective motion, in
addition to the thermal one. Electrons occupy a set of quantized Landau
levels in the transverse plane, while their distribution
parallel to the field is taken to be a 1-D
relativistic Maxwellian at temperature $T_e$,
superimposed to a bulk motion with velocity $\beta_{bulk}$ (in units
of the light velocity $c$). In turn, both  $T_e$ and $\beta_{bulk}$ are
treated as free model parameters.
We then built two versions of the code. The first one is
computationally faster, and scattering is
treated in the magnetic Thomson limit, neglecting electron recoil along
the field direction. This means that simulated spectra are only
valid up to a few tens of keV ($h \nu <mc^2/\gamma$~keV,
$B/B_{QED} < 10$). The second version of the code is computationally
heavier, but it includes the relativistic QED resonant cross section   
\citep[see][and references therein]{ntz2},
which is required to extend spectral predictions in the hard X-rays band
(i.e. up to and beyond the \INT\ energy range).

\section{Predicted Spectral and polarization properties}

\subsection{Results computed in the Thomson limit}

\begin{figure}
\begin{center}
\includegraphics*[width=10cm]{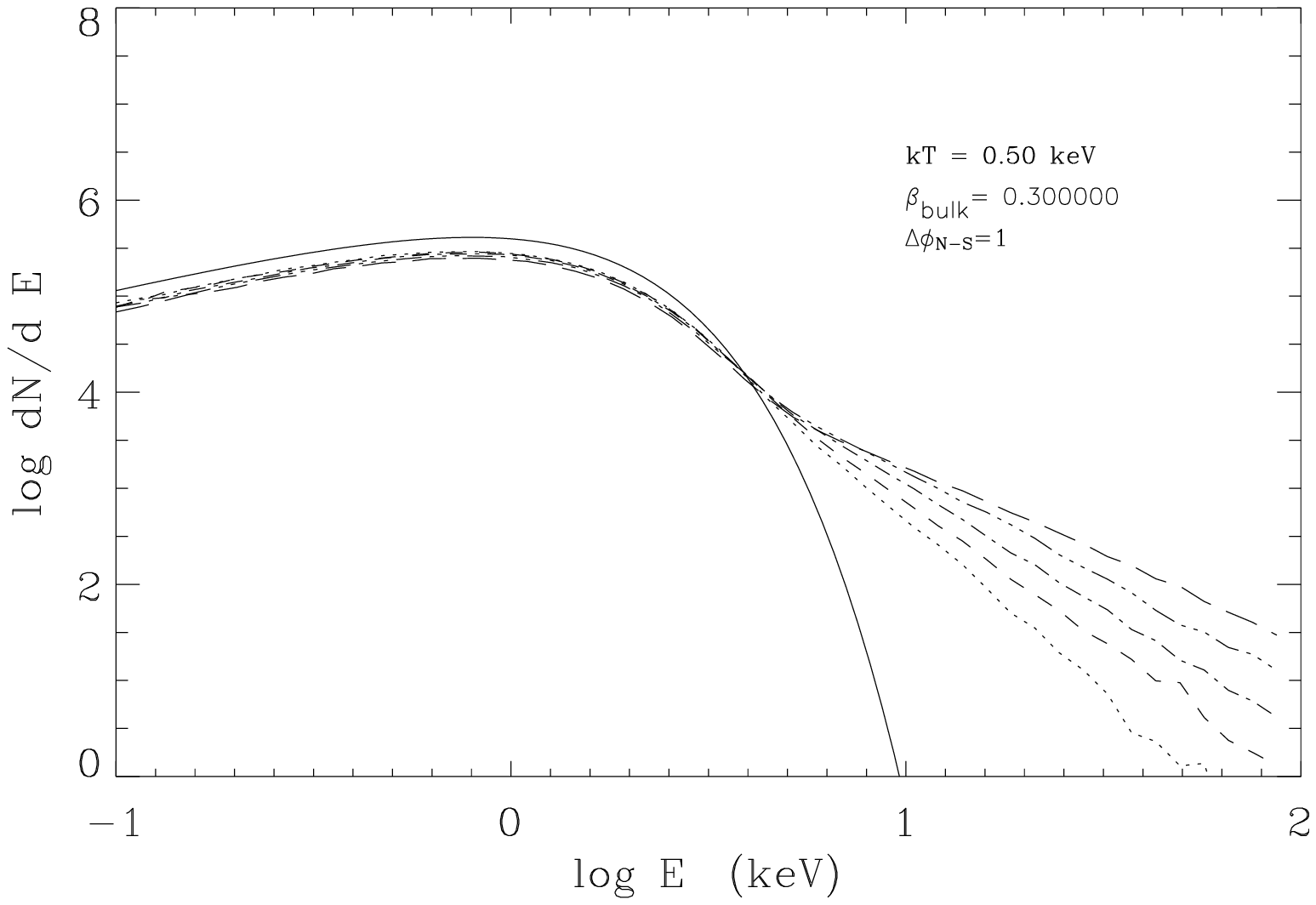}
\includegraphics*[width=10cm]{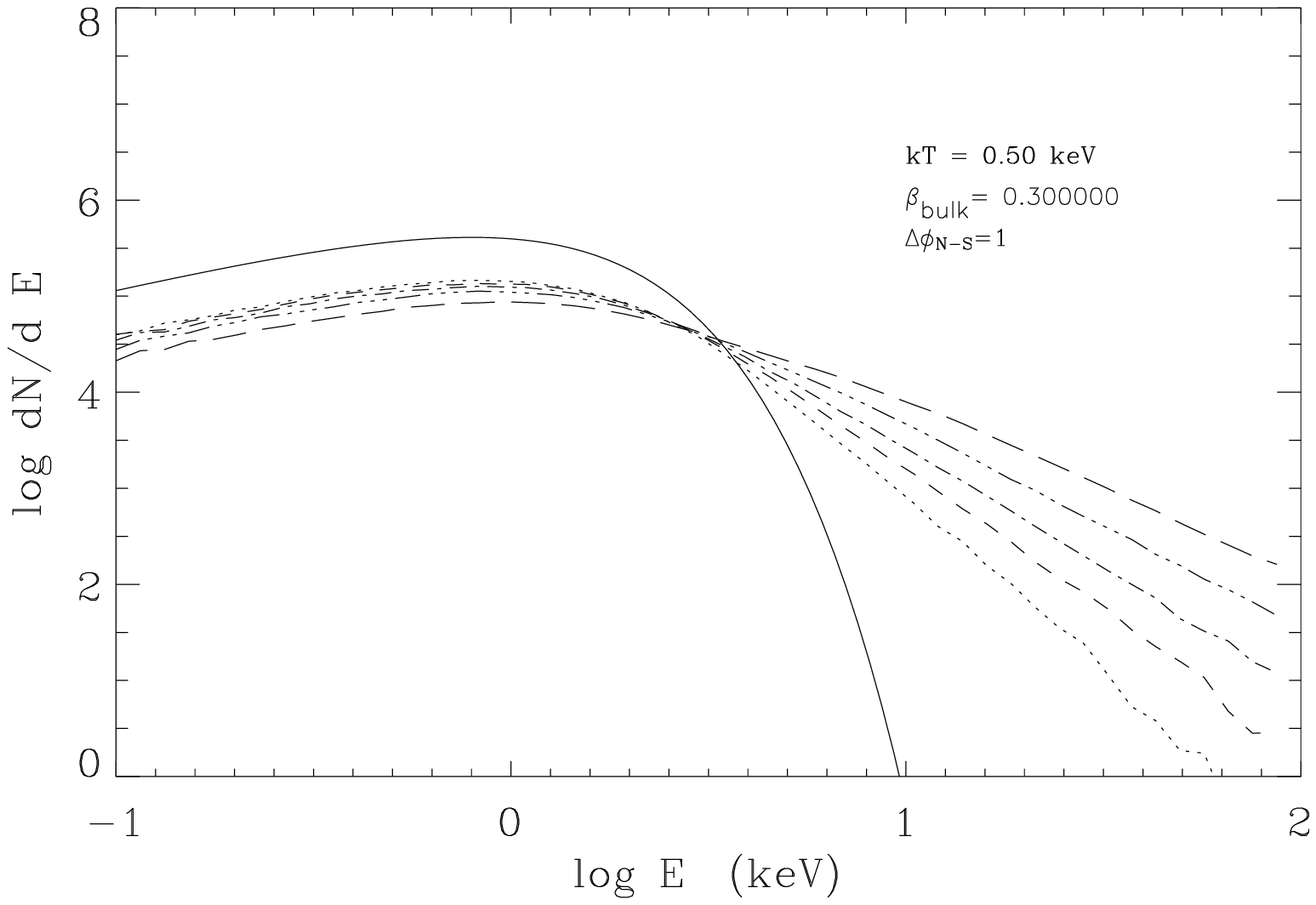}
\end{center}
\caption{Computed spectra for $B = 10^{14}~G$, $kT  = 0.5$~keV,
$\beta_{bulk} =
0.3$, $\Delta \phi = 1$ and different values of $kT_e$: 5 keV (dotted),
15 keV (short dashed), 30 ~keV (dash-dotted), 60 ~keV (dash-triple dotted)
and 120 keV
(long dashed).
The solid line represents the seed
blackbody.
The two panels
correspond to two different values of the magnetic colatitude of the
observer:
$\Theta_s=64^\circ$ (top) and $\Theta_s=116^\circ$ (bottom).
Figure from \citet{ntz1}; see the original paper for all 
details.
}
\label{f1}
\end{figure}

\begin{figure}
\begin{center}
\includegraphics*[width=10cm]{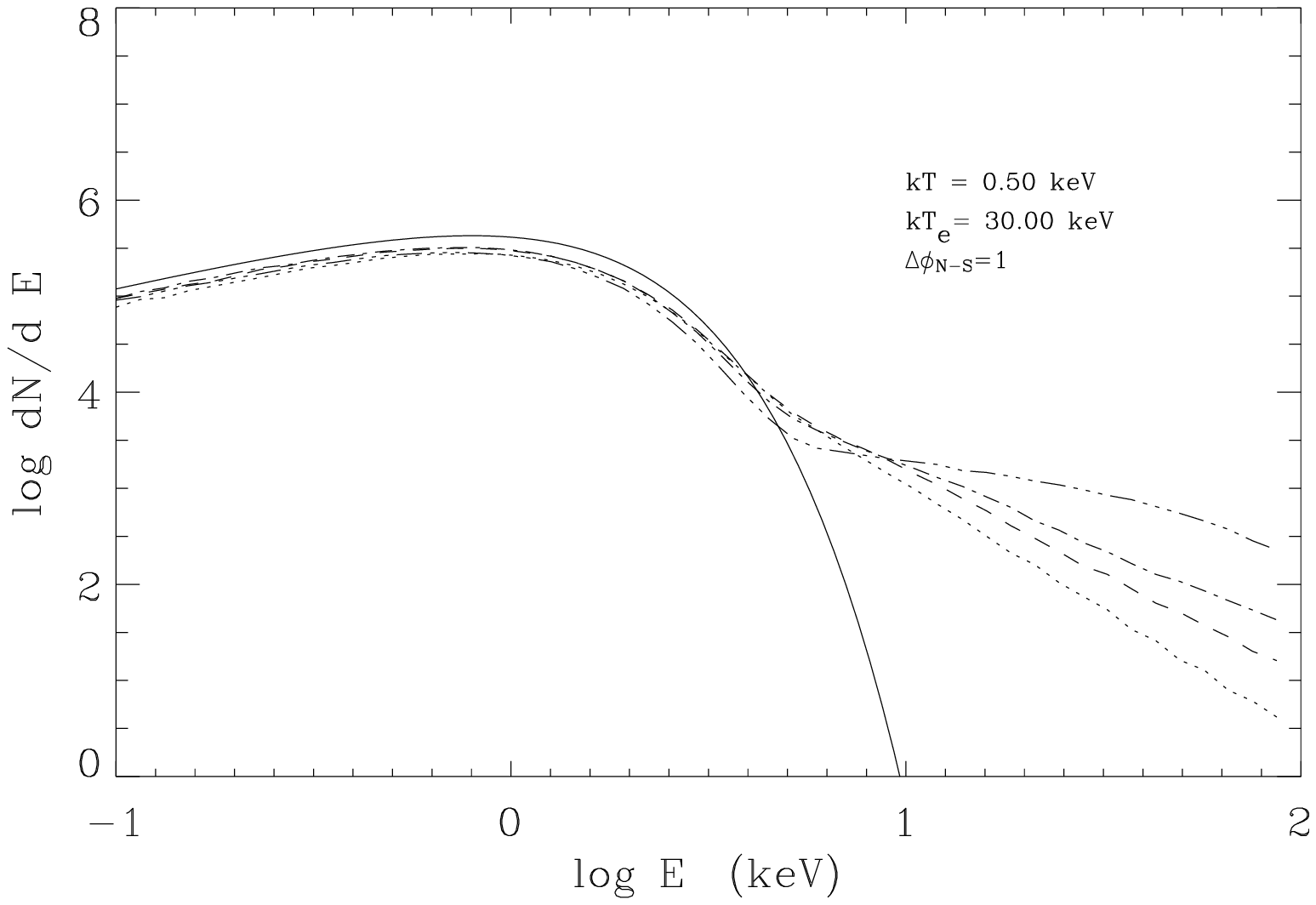}
\includegraphics*[width=10cm]{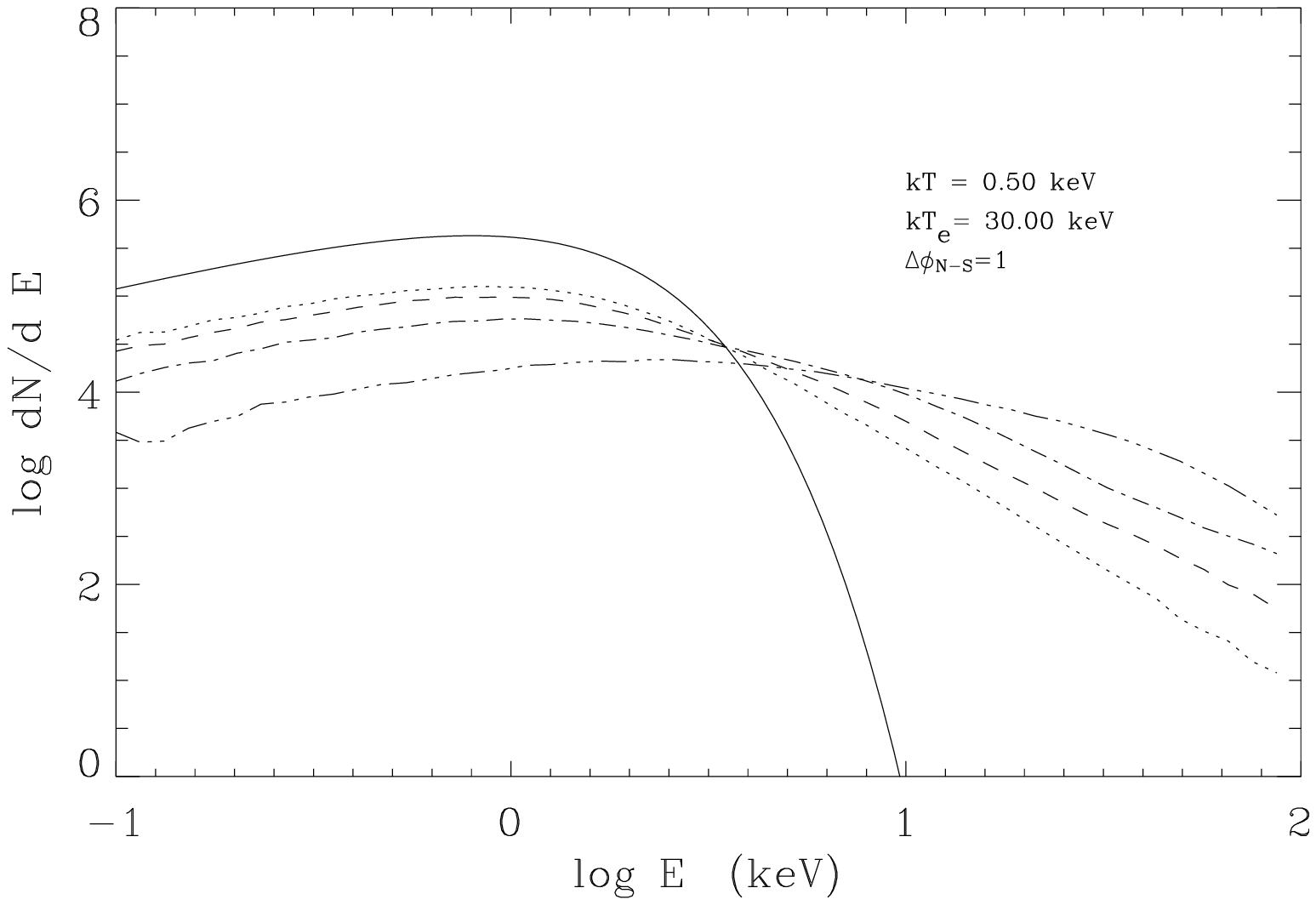}
\end{center}
\caption{
 Computed spectra for $B = 10^{14}~G$, $kT  = 0.5$~keV, $kT_e =
30$~keV, $\Delta \phi = 1$ and different
values of $\beta_{bulk}$: $0.3 $ (dotted),
$0.5 $ (short dashed), $0.7 $ (dash-dotted)
and $0.9$ (dash-triple dotted).   
The solid line represents the seed
blackbody.
The two panels
correspond to two different values of the magnetic colatitude of the
observer:
$\Theta_s=64^\circ$ (left) and $\Theta_s=116^\circ$ (right).
Figure from \citet{ntz1}; see the original paper for all 
details.
}
\label{f2}
\end{figure}

Figs.~\ref{f1}, \ref{f2} show a series of spectra computed in
the Thomson limit, and illustrate
the effects on the spectral
shape of varying $\beta_{bulk}$ and
$kT_e$,  respectively.
As mentioned above, although the curves in these figures
extend up to $\sim 1000$~keV, being these spectra computed in
the non-relativistic limit they should only be considered valid
up to a few tens of keV.
Spectra are  plotted for two values
of the magnetic colatitude at infinity,
$\Theta_s$,
one for each hemisphere (top and bottom panels).
By comparing the left and right panel in each figure, we can immediately
notice the absence of symmetry between the north and the south
hemispheres: as $\Theta_s$ increases, spectra become more and more  
comptonized. This
reflects the fact that in our model,
which accounts for the charges bulk velocity, currents
flow has a preferential direction (in this simulation from the north to
the south pole along the field
lines). As a consequence,
an observer located in the northern hemisphere ``sees''
currents flowing away from him and observes spectra that are 
less comptonized with respect to an
observer in the southern hemisphere (which "sees" currents flowing towards
him). Of course the opposite choice for the current direction would
simply result in $\Theta_s \to 180^\circ - \Theta_s$.

As it can be seen from Figs.~\ref{f1}, \ref{f2}, an
increase in either $\beta_{bulk}$ or $kT_e$ always corresponds to an
increase in the comptonization degree of the spectrum. The same is true 
for an increase in the twist angle, $\Delta \phi$, which translates in an
increase in the electron number density.  As shown in the 
bottom panel of Fig.~\ref{f2}, the effect is
particularly
notable in the case of $\beta_{bulk}$.
If $\beta_{bulk} >  0.5$, an observer located in the southern
hemisphere (i.e. with currents flowing towards him)  sees a spectrum   
which is no more peaked at $\sim kT$, but peaks instead at
about the thermal energy of the scattering
particles.
This is because under these conditions Comptonization starts to saturate 
and
photons fills the Wien peak of
the Bose-Einstein distribution. This is observationally important, since
it means that in some cases the observed thermal bump may be
unrelated to the peak of the seed blackbody, but instead may give  a
direct information on the energy of the magnetospheric particles.
Moreover, for intermediate values of the parameters,  spectra can
be double humped, with a downturn between the two humps.

\begin{figure}
\begin{center}
\includegraphics*[width=10cm]{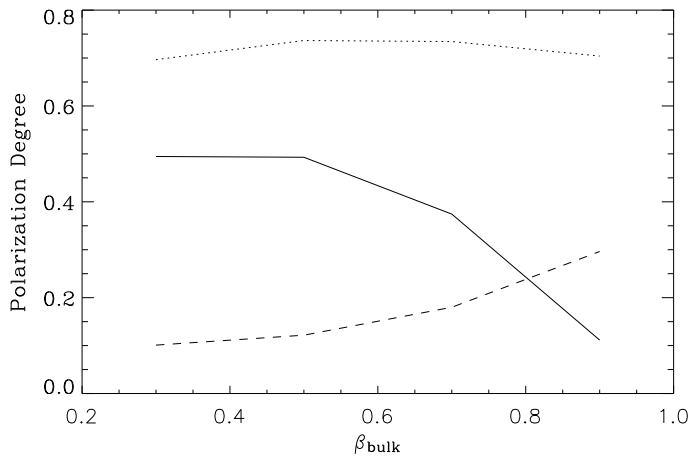}
\end{center}
\caption{Degree of polarization ($0.1-1000$~keV band) as a function of 
$\beta_{bulk}$ for
$kT_e = 30$~keV and $\Delta \phi = 1$, $B=10^{14}$~G, $kT = 0.5$~keV,
$\beta_{bulk} = 0.3$. Different curves correspond to: seed photons 100\%
polarized
in the ordinary (solid line),  extraordinary mode (dotted line), and
unpolarized (dashed line). 
Figure from \citet{ntz1}; see the original paper for all 
details.
} 
\label{polla}

\end{figure}

\subsection{Looking at magnetars with polaroid glasses}

In Fig.~\ref{polla} we show, as a function of $\beta_{bulk}$, the degree
of polarization of the emerging radiation,
defined as $\vert N_{extr} - N_{ord}\vert/(N_{extr} + N_{ord})$ where
$N_{extr}$ and $N_{ord}$ are, respectively, the number of ordinary and
extraordinary photons collected at infinity. The polarization degree is 
computed in the $0.1-1000$~keV band, it has
been averaged over frequency, over the whole emitting surface and over the
sky at infinity. As it can be seen, the efficiency at
which completely polarized surface radiation
is depolarized increases by increasing the strength
of magnetospheric upscattering, i.e. by increasing
$\beta_{bulk}$. We found similar results when increasing $kT_e$
or $\Delta \phi$. The depolarization effect
is stronger for ordinary seed photons, for which the probability
of undergoing mode
switching in the scattering process is higher.
As the figure shows, would the
surface radiation be completely unpolarized, while
passing through the magnetosphere,  it can
acquire only a relatively small degree of linear polarization: typically 
10-20\%, up to 30\% for very extreme values of the current  bulk
velocity.
This means that, would future observations of X-ray polarization result in
measurements larger than 10-30\%, the excess has to be attributed to an 
intrinsic property of the surface radiation.

\subsection{Relativistic, QED scattering}

As previously mentioned, the treatment based on the Thomson limit becomes
inadeguate when modelling the emission at higher energies, as that
observed by \INT\ up to $\sim 200$~keV. Proper
investigation of this range  demands a complete QED treatment of magnetic
Compton scattering. This is mandatory independently on the energy of the
scatterers.  If  highly
relativistic particles are considered, a photon can be boosted to
quite large energies in a single scattering (making electron recoil
important) and, if it propagates towards the star, it may scatter
again where the field is above the QED limit. Even in presence of mildly
relativistic particles, in order to populate the hard tail it is necessary
that soft photons experience multiple scatterings, during which the
resonant condition is matched in regions characterized by a progressively
higher field, again making  QED effects non-negligible.
 
The Compton cross-section for electron   
scattering in the presence of a magnetic field was first studied in the
non-relativistic limit by \citet{clr71}, and the QED
expression was derived long ago by many authors
\citep{her79,dh86,bam86,hd91}.
However, its form is so complicated to be often of little practical
use in numerical calculations. On the other hand, since we are only
interested in making spectral predictions for the continuum emission (and
not on the details of the cyclotron lines profile), in our case we can
safely assume that scattering occurs {\it at resonance}. We recently   
investigated this limit, and presented a complete, workable set of 
expressions for the QED cross section which can be used in our Monte
Carlo simulations of photon scattering \citep{ntz2}. Our expressions are
completely general in the sense that no limiting assumption is made on the
magnetic field strength, and can be generalized to scattering
onto positive charges (positrons). However, our present aim is to use them
for
spectral modelling in the $\,\sim 0.1$-200 keV range, in which case
resonant scattering of soft photons necessarily occurs where $B/
B_{QED} <
1$. This allows us to restrict to the case in which
Landau-Raman scattering occurs only up to the second Landau level, which
makes the numerical computation faster. Photon spawning during the
de-excitation is consistently accounted for.
In order to verify the necessity of the QED computation,  we compared our
cross sections with the non-relativistic limit, and we found
significant deviations even for
$B/B_{QED} > 0.1$.

\begin{figure}
\begin{center}
\includegraphics*[width=10cm]{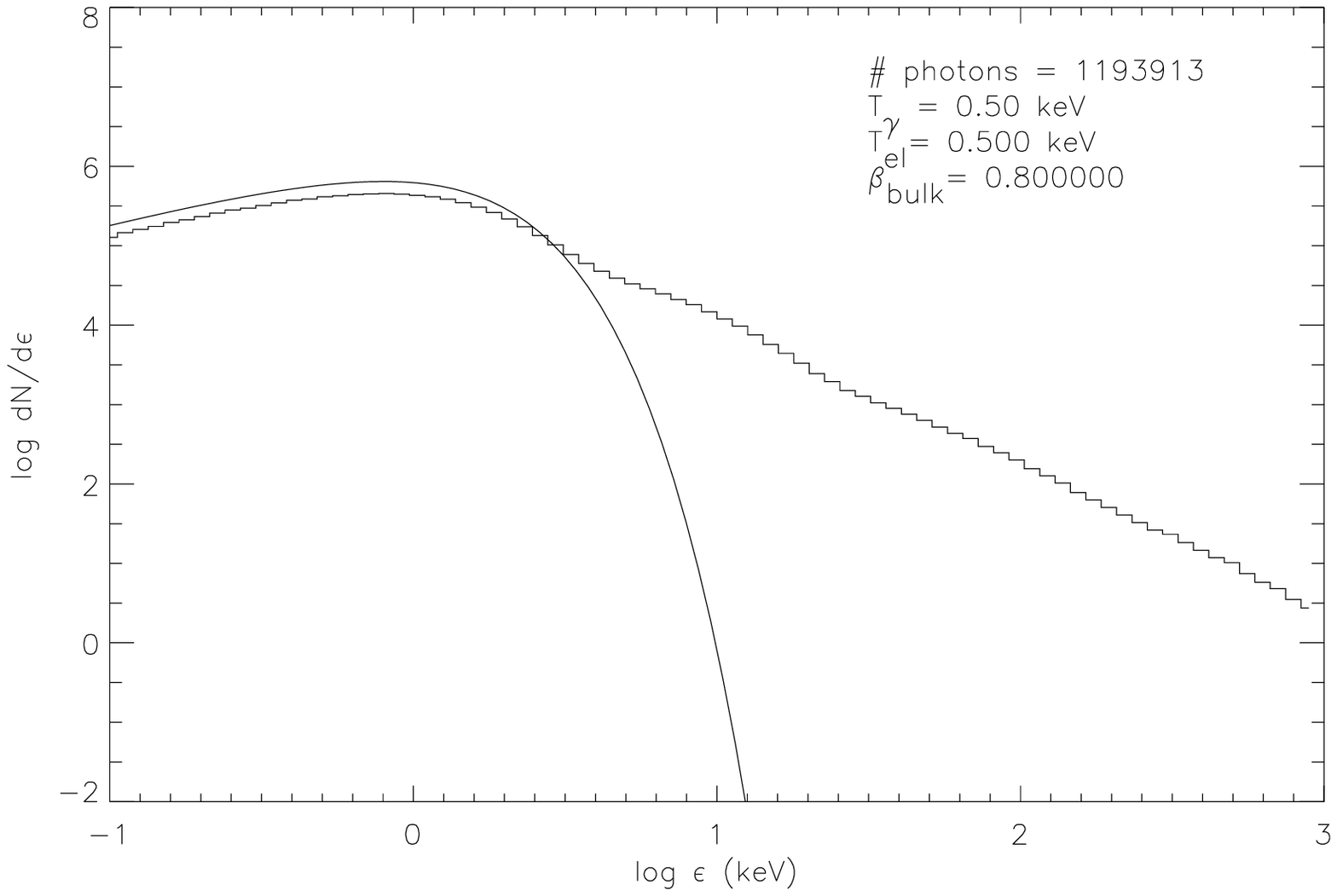}
\includegraphics*[width=10cm]{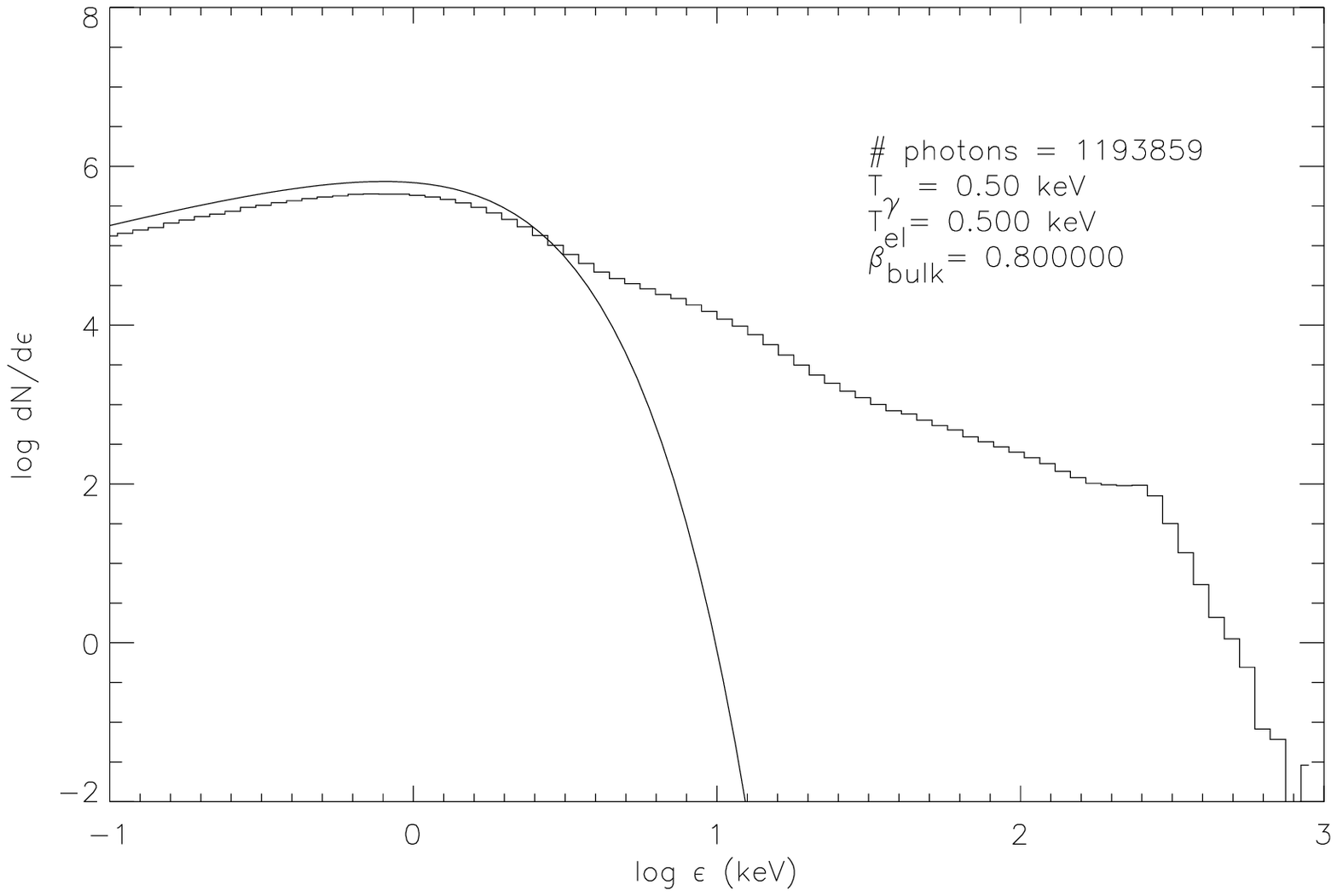}
\end{center}
  \caption{
 Computed spectra for $B = 10^{14}~G$ and
$\Delta \phi = 1$. The top panel shows spectra computed by using the
cross section in the Thomson limit, while on the bottom we use the full 
QED
expression of resonant scattering. Here electrons are mildly relativistic
($\gamma = 1.7$).
}
\label{f3}  
\end{figure}

\begin{figure}
\begin{center}
\includegraphics*[width=8cm]{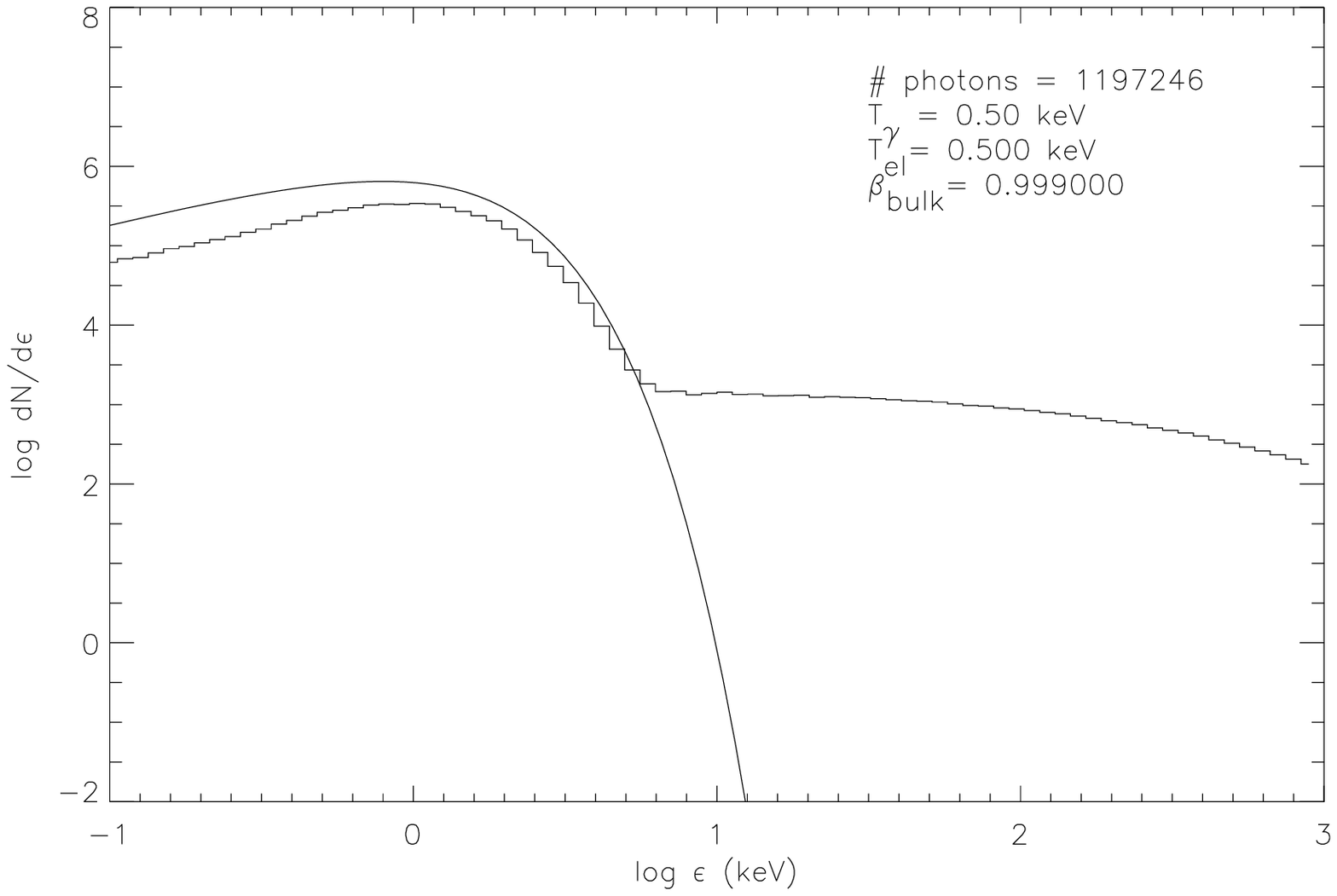}
\includegraphics*[width=8cm]{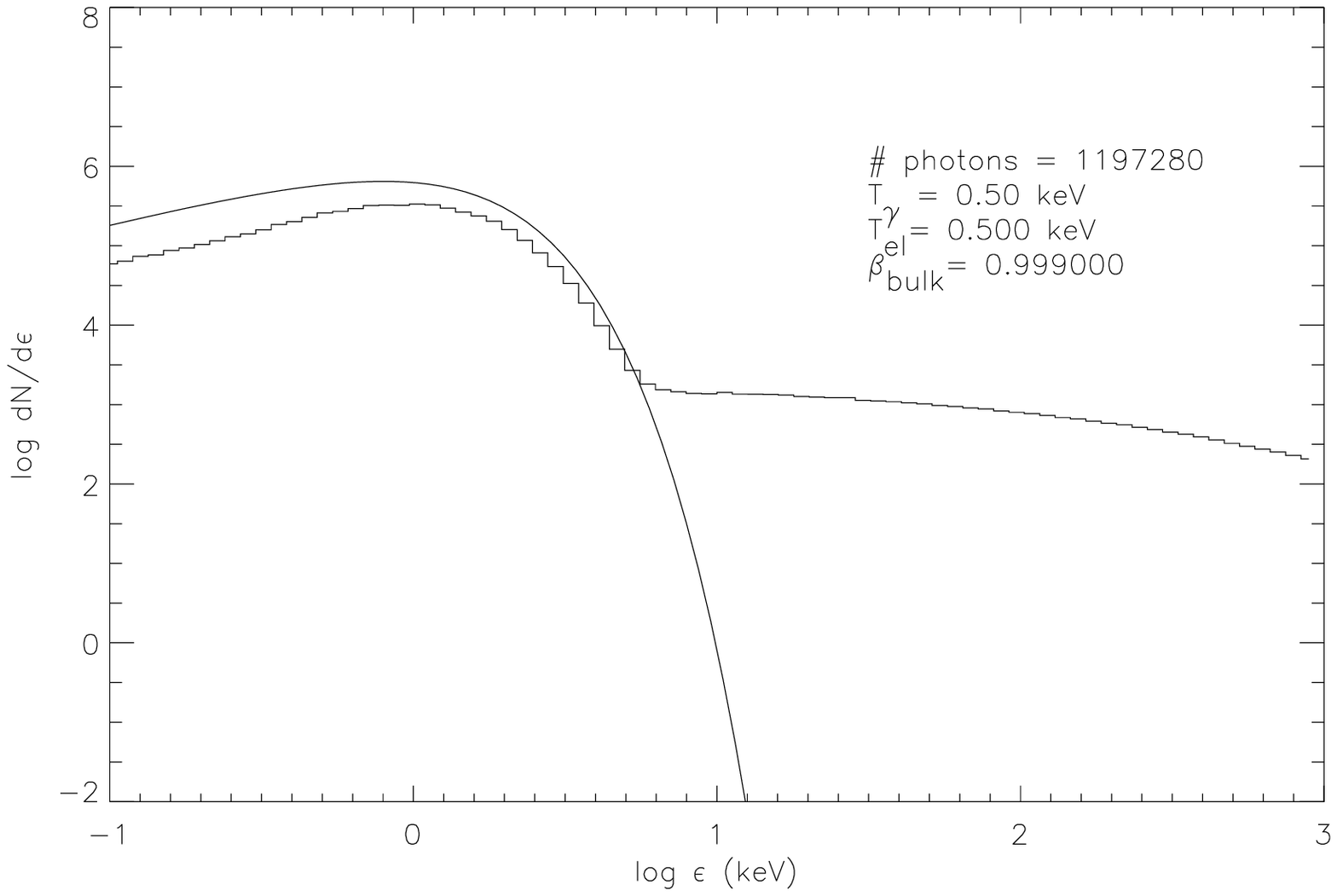}
\end{center}
  \caption{ Same as in Fig.\ref{f4}, but for highly relativistic electrons 
with $\gamma = 22$.
}
\label{f4}  
\end{figure}

In Figs.~\ref{f3},\ref{f4} we show a few spectra computed by using our
Monte Carlo code with the
full
QED scattering cross section (top panel), and their analogues computed
by using the cross section
in the Thomson limit (bottom). As we can see, below a few tens of keV the
two spectra are identical, confirming the validity of the use of the
Thomson limit when making predictions in the \XMM\ range. On the other
hand,
if electrons are mildly
relativistic (Fig.~\ref{f3}), when considering self-consistently electron
recoil and QED effects the spectrum exhibits a break at higher energies. 
This is due to the fact that, if the Lorentz factor of electrons is
$\gamma \sim$ a  few, each scattering characterized by a limited  energy 
gain. Therefore, in order to populate the hard energy tail
soft seed photons need to experience a series of successive
scatterings. On the other hand, the efficiency of the QED cross section 
decreases with
increasing energy (or, being the process resonant, with increasing the
magnetic field) and the combination of these two effects leads to the 
appearance of the spectral break. Unfortunately, this means that the
energy of the break depends on a series of processes and on the details of
the magnetic field configuration and its currents distribution and
therefore can not be predicted a priori nor estimated using a simple
expression. Further work on this issue is in preparation.
On the contrary, if the electrons are more relativistic, the energy   
gain per scattering is much larger and the hard tail becomes efficiently
populated after just a few scatterings. Therefore, independently on the
details of the cross section, in this case we do expect the formation of a
spectral tail unbroken, 
even up to $>1000$~keV (see Fig.~\ref{f4}).

\section{Fitting magnetars spectra in the soft X-ray range}

\begin{figure}
  \includegraphics[height=.3\textheight,angle=270]{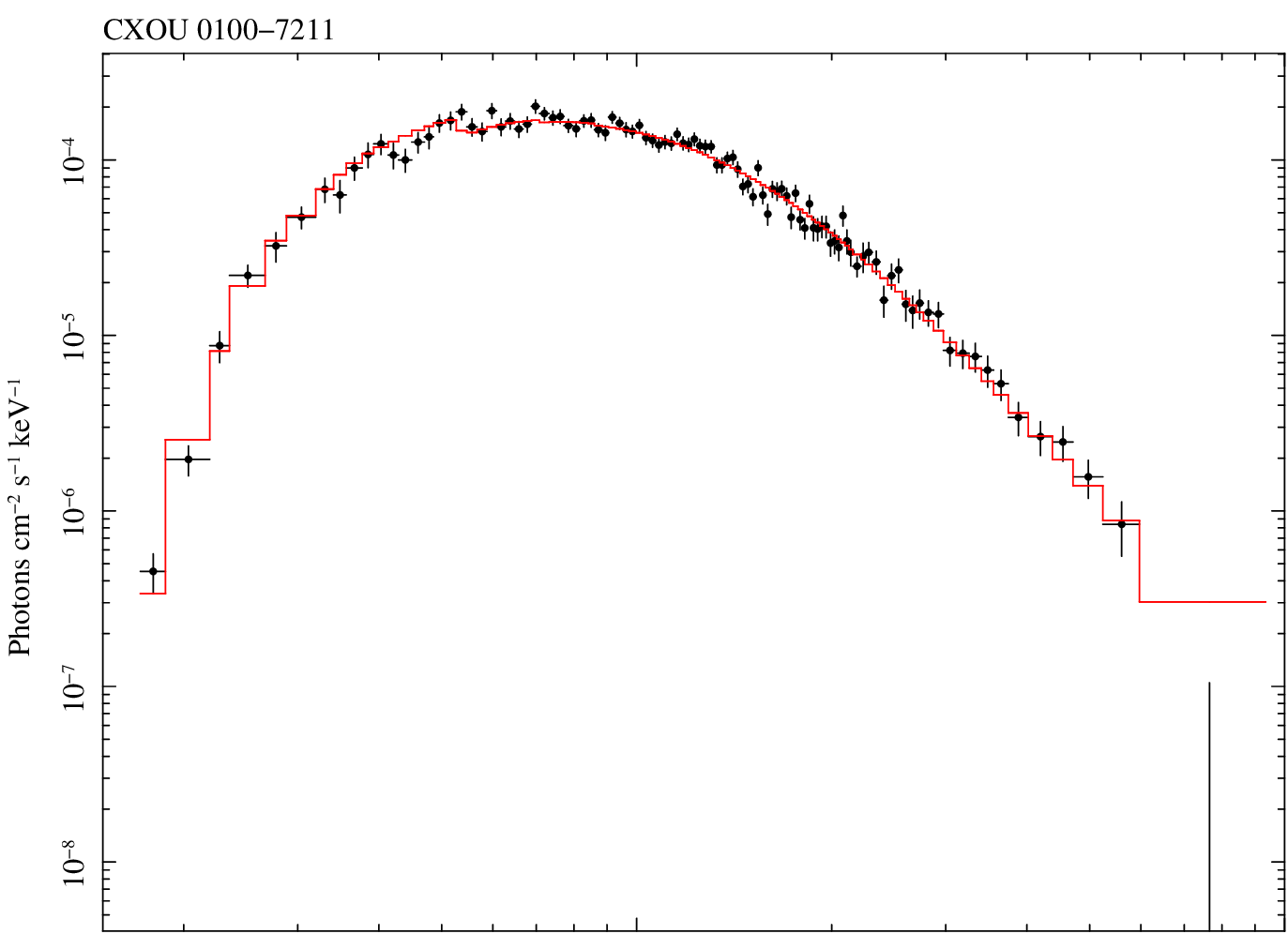}
  \includegraphics[height=.3\textheight,angle=270]{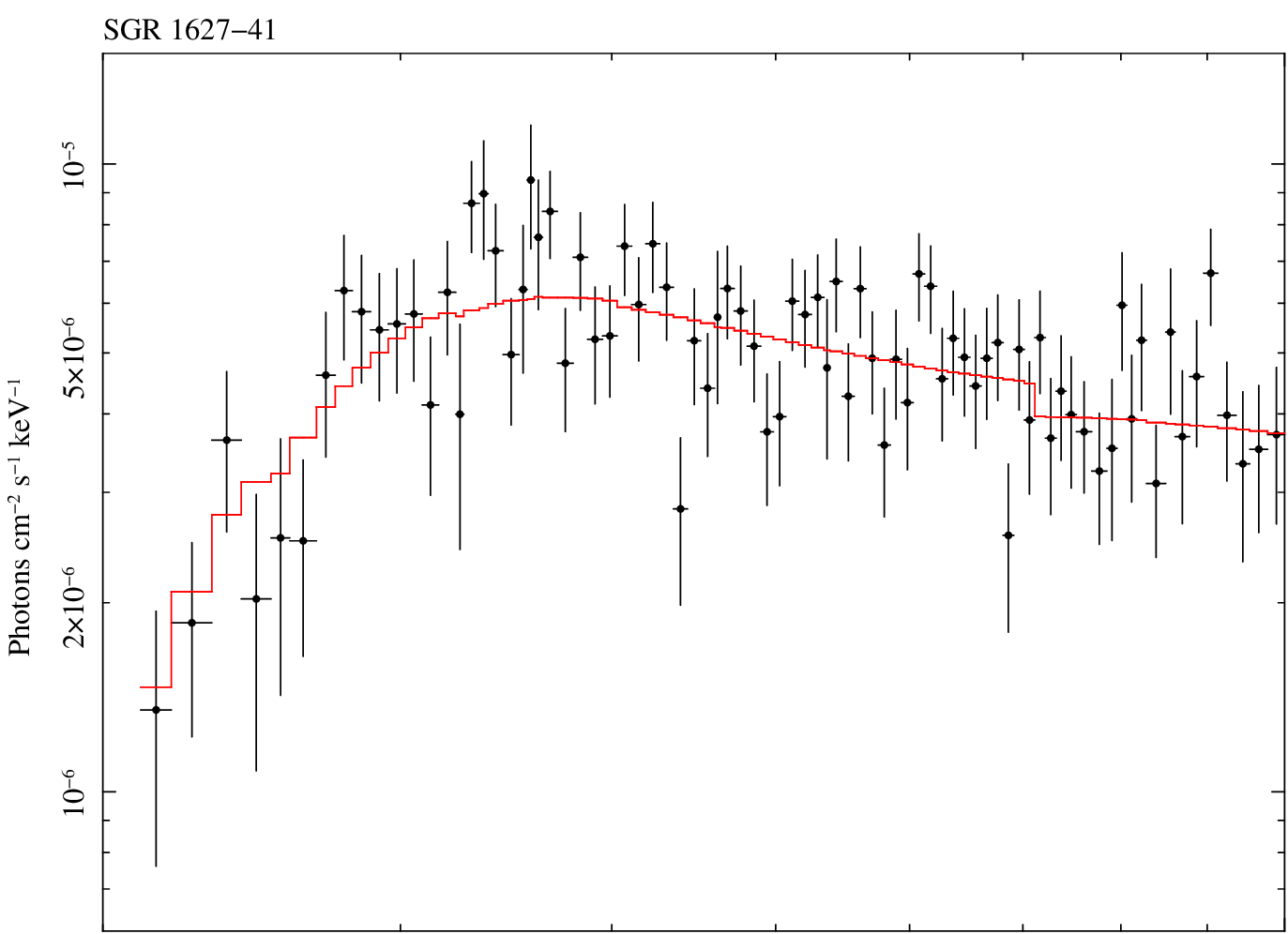}
\end{figure}

\begin{figure} 
\includegraphics[height=.3\textheight,angle=270]{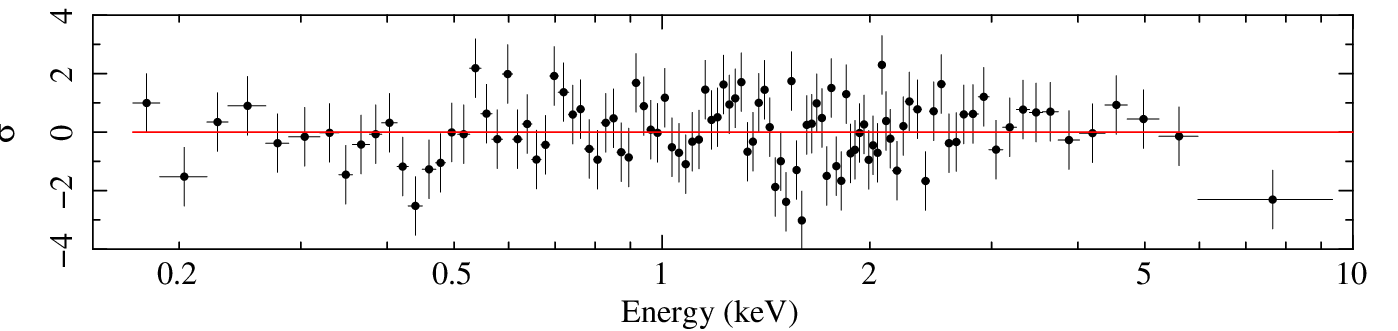}
  \includegraphics[height=.3\textheight,angle=270]{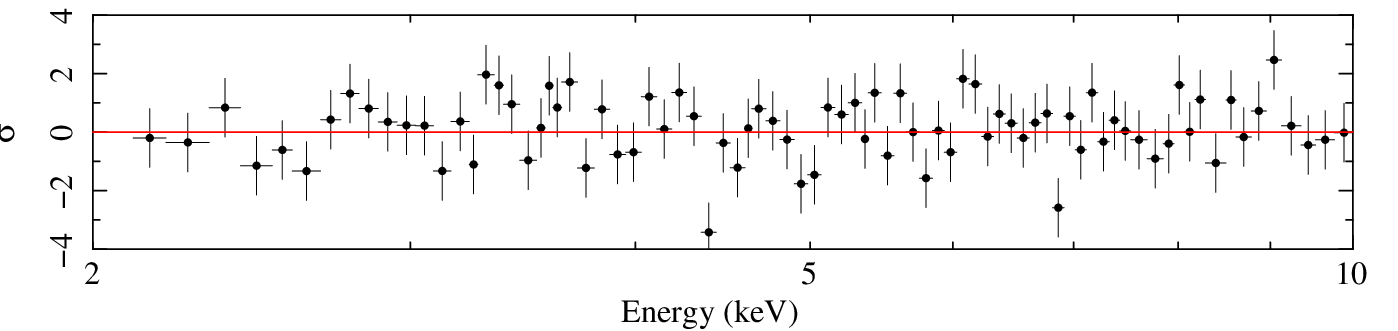}
  \caption{
Fit of the \XMM\ spectra of \cxo\ and \sgrd\ with the NTZ 
model.
The
fitting has been restricted to the
$1-10$~keV range \cite[figure re-adapted
from][see details therein]{papntz}.
}
\label{figfit1}
\end{figure}

\begin{figure}
\includegraphics[height=.2\textheight,angle=270]{1e1547_twist_ufs.ps}
  \includegraphics[height=.2\textheight,angle=270]{1e1048_twist_ufs.ps}
  \includegraphics[height=.2\textheight,angle=270]{sgr1806_twist_ufs.ps}
\end{figure}  
  
\begin{figure}
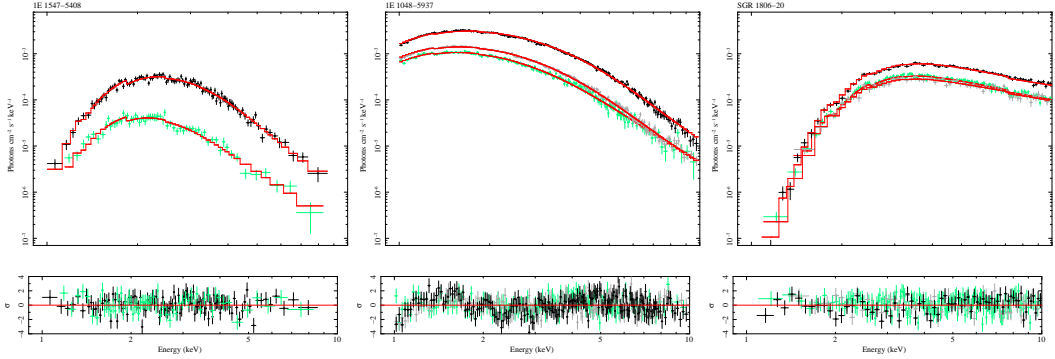
 
\includegraphics[height=.2\textheight,angle=270]{1e1547_twist_res.ps}
  \includegraphics[height=.2\textheight,angle=270]{1e1048_twist_res.ps}
  \includegraphics[height=.2\textheight,angle=270]{sgr1806_twist_res.ps}
  \caption{   
Same as in Fig.~\ref{figfit1} for
the AXPs \1e, \ee\ and for  \sgra. In the case of these sources,
we  performed a joint
fit of spectra taken at three different epochs \cite[figure re-adapted
from][see details therein]{papntz}.
}
\label{figfit2}
\end{figure}

Our non-relativistic code has been used to produce
an archive of spectral models
that have been subsequently implemented in XSPEC (NTZ model). No attempt  
is made to
fit the value of the
polar field strength, that has been fixed at $10^{14}$~G, and, in order to
miminise the number of free parameters, the models in the archive were
computed by assuming that  the electron temperature is related to  
$\beta_{bulk}$ via equipartition \citep[see][for details]{ntz1}.
We recently applied this model to a large sample of magnetars spectra  
taken in quiescence. We also considered magnetars that exhibit
long-term spectral variation, in which cases we
considered a set of two-three
observations for each source, corresponding to different spectral states,
and performed a joint fit by assuming that only the value of the
interstellar absorption remains fixed. A few examples are shown in
Figs.\ref{figfit1},\ref{figfit2} and, for all
details
and best fit parameters, we refer the reader to \citet{papntz}. Our main
result is that, when we restrict to the 1-10~keV band (by using \XMM\  
data) the NTZ model successfully reproduces the soft X-ray part of
the spectrum of most of the sources (apart from \ea\ and \uu, which
are
discussed separately in \citet{papntz}), as well as the long
term spectral
variation observed in a few sources, without the
need of additional components.  This represents a substantial improvement
with respect to previous attempts to model magnetars quiescent emission in
the same energy band with a simpler 1-D RCS model \citep{nanda2}, where 
it
was found that in a few cases a PL component was required, in addition to
the RCS one, to provide an acceptable fit to the data below 10~keV.
We also fitted the combined \XMM\ and \INT\ spectra of sources with
NTZ+PL model, to
investigate whether a NTZ model can account for the soft X-ray component
in the full spectral energy distribution. Since our XSPEC model is
restricted to Thomson limit, for self-consistency it has been
artificially truncated at
$10$~keV and the additional PL is required to simulate the observed hard
tail. Also in this case, we have been able to find a successful fit.
However, the fit converges
with a hard X-ray component that gives a substantial contribution when
extrapolated to the soft X-ray band and, consequently, the best fitting
parameters of the NTZ model are substantially different from those we
found fitting the 1-10~keV emission only. Whether it is possible that the
extrapolation of our spectrum into the hard X-ray domain, when computed
by self-consistently accounting for QED effects, is responsible also for
the observed \INT\ tail is matter of future investigations.

\section{Conclusions}
\label{conc}

Results presented here probe that the twisted magnetosphere model, within
the magnetar
scenario, is in general agreement with observations.
As discussed, our 3-D model of resonant scattering
of thermal, surface photons reproduces several AXPs and SGRs spectra
below 10~keV with no need of extra components. Simulated spectra computed
accounting for the whole QED cross sections show that, if magnetospheric 
electrons are mildly relativistic, a spectral break appears at
higher energies. The energy of the spectral break, however, is strongly
dependent on the model details.

Our simulations are based on a number of assumptions, among which is the
fact that the magnetic field is a simple twisted dipole.
We recently investigated the solution
for twisted fields with more general force-free configurations,
including multipolar component \citep[see][]{lucia} which allow 
us to
simulate scenarios in which the twist is localized. Nevertheless, we 
caveat that, at the present status of art, is the
physical
structure of the magnetosphere which is still an open problem. All our
fits
require mildly relativistic electrons ($\gamma \sim 1$), therefore the
resonant scattering model works
well provided that a braking mechanism exists to
maintain electrons in a non-relativistic regime \cite[see][this issue]
{luciano}. 

Further works aimed at testing the RCS  emission
scenario against the variability that is shown during the outbursts of
some transient AXPs are in preparation (Albano et al., in prep.,  
Bernardini et al., in prep., Israel et al., in prep.).  Also, we plan to 
use 
our QED code to study spectral formation and
polarization
pattern up to $\sim 200$~keV and to investigate the emission mechanism
responsible for the hard X-ray tails detected by \INT\
(photon upscattering,
curvature radiation from particles in the external magnetosphere, or
other mechanisms).
These results will be presented in forthcoming papers.

\section{Acknowledgments}

RT and LN are partially
supported by INAF-ASI through grant AAE-I/088/06/0. NR is
supported by an NWO Veni Fellowship.
We thank A. Tiengo and P. Esposito for the data of \cxo\ and \sgrd,

\end{document}